\documentstyle[12pt,epsfig]{article}
\def\gtrsim{\ \lower -2.5pt\hbox{$>$} \hskip-8pt \lower 2.5pt 
\hbox{$\sim$}\ }
\def\lesssim{\ \lower -2.5pt\hbox{$<$} \hskip-8pt \lower 2.5pt 
\hbox{$\sim$}\ }

\begin{document}
\hspace{10.5cm}{UB/FI 96-1}

\hspace{10.5 cm}{February 1996}

\vspace{5 mm}
\centerline{\bf {\LARGE A complete solution to neutrino mixing}}
\vspace{15 mm}

\centerline{G. Conforto, A. Marchionni, F. Martelli, F. Vetrano}
\centerline{\em Universit\`a degli Studi, 61029 Urbino, Italy}
\centerline{\em Istituto Nazionale di Fisica Nucleare, 50125 Firenze, Italy}
\vspace{5 mm}
\centerline{M. Lanfranchi}
\centerline{\em Dipartimento di Fisica dell'Universit\`a and}
\centerline{\em Istituto Nazionale di Fisica Nucleare, 50125 Firenze, Italy} 
\vspace{5 mm}
\centerline{G. Torricelli-Ciamponi}
\centerline{\em Osservatorio Astrofisico di Arcetri, 50125 Firenze, Italy}
\vspace{8 mm}
\centerline{\em (To be published in Astroparticle Physics)}
\vspace{8 mm}
\centerline{\bf Abstract}
Deviations from expectations have been claimed for solar, 
atmospheric and high energy prompt neutrinos from charm decay.
This information, supplemented only by the existing very good 
upper limits for oscillations of the $\nu_{\mu}$ at accelerator 
energies, is used as input to a phenomenological three-flavour 
analysis of neutrino mixing. The solution found is unique and 
completely determines the mass eigenstates as well as the 
mixing matrix relating  mass and flavour eigenstates. Assuming 
the mass eigenstates to follow the hierarchy $m_{1} \ll 
m_{2} \ll m_{3}$, their values are found to 
be $m_{1} \ll 10^{-2}$ eV, $m_{2} = (0.18 \pm 0.06)$ eV,
$m_{3} = (19.4 \pm 0.7)$ eV. 
These masses are in agreement with the leptonic quadratic 
hierarchy of the see-saw model and large enough to render 
energy-independent any oscillation-induced phenomenon in solar 
neutrino physics observable on Earth. This possibility is not 
excluded by the present knowledge of solar neutrino 
physics. The mixing angles are determined to be 
$\theta_{12} = 0.55 \pm 0.08$, $\theta_{13} = 0.38 \pm 0.06$, 
$\theta_{23} < 0.03$. Small values of $\theta_{23}$ are typical 
of any solution in which $m_{3}$ lies in the cosmological 
interesting region. The solution found is not in serious 
contradiction with any of the present limits to the existence 
of neutrino oscillations. The most relevant implications in 
particle physics, astrophysics and cosmology are discussed.

\footnotetext {e-mail addresses:\newline 
{\em conforto@vxcern.cern.ch, marchion@vxcern.cern.ch, 
martelli@vxcern.cern.ch, \newline vetrano@fis.uniurb.it, 
lanfranchi@fi.infn.it,torricel@arcetri.astro.it}}

\section{Introduction}

In recent years, several deviations from expectations 
have been reported in various fields of neutrino physics 
\cite{JSch93}.

At high energy, the equality between the $\nu_{e}$ and 
$\nu_{\mu}$ energy spectra prescribed by e-$\mu$ universality 
for the semi-leptonic decays of charmed particles appears to be 
violated at the four sigma level.

In atmospheric neutrinos, the ratio between the $\nu_{\mu}$ and 
$\nu_{e}$ fluxes is found to be substantially less than the 
predicted value of 2.
 
Even if quantitative conclusions depend on the somewhat different 
predictions of the various solar models, measurements of solar 
neutrino fluxes on Earth have been known for some time to fall 
short of the corresponding theoretical expectations. 

These positive results have been taken as indications of the 
existence of non-zero neutrino masses and mixings. However, they 
have to be examined for consistency with each other and confronted 
with the many negative searches for oscillations and their 
corresponding upper limits. 

No satisfactory purely phenomenological over-all interpretation 
has been obtained so far. 

In this paper we describe a comprehensive three-flavour analysis 
which results in an unique, overconstrained and complete description 
of neutrino mixing. Negative results being usually available only 
at the 1.64 sigma level (90\% CL), we have used as input data only 
the three positive results above and the two most stringent upper 
limits obtained in the searches for $\nu_{e}$-$\nu_{\mu}$ and 
$\nu_{\mu}$-$\nu_{\tau}$ oscillations at high energy accelerators. 
We also discuss the compatibility of this solution with the other 
existing limits and its most relevant implications in particle
physics and astrophysics.

Preliminary accounts of this work have been presented elsewhere 
\cite{FMar94,GCon95}.

CP-invariance is assumed throughout the paper. Also, unless 
otherwise stated, the term neutrino is used to indicate both 
neutrino and antineutrino.

\section{Three-flavour formalism}

In the complete three-flavour approach, the weak 
eigenstates $|\nu_{\alpha}\rangle = \nu_{e}, \nu_{\mu}, \nu_{\tau}$ 
and the mass eigenstates $|\nu_{i}\rangle = \nu_{1}, \nu_{2}, \nu_{3}$ 
are related by an unitary transformation matrix $U$, in terms of 
which the probability of an initial neutrino $\nu_{\alpha}$ 
of energy $E$ being equal to another neutrino $\nu_{\beta}$  
at a distance $L$, can be written as
\begin{equation}
P_{\alpha\beta} = \delta_{\alpha\beta} - 4 \sum_{j>i} U_{\alpha i} 
U_{\beta i}U_{\alpha j}U_{\beta j} \sin^{2}(\Delta_{ij}/2)
\label{eq:1}
\end{equation}
with $\Delta_{ij} = \delta m_{ij}^{2} L/2E$, where 
$\delta m_{ij}^{2} = m_{i}^{2} - m_{j}^{2}, ~m_{i} = m(\nu_{i})$. 

The $U$-matrix can be parametrized as
\begin{equation}
U = \left( \begin{array}{ccccc}
c_{12}c_{13} && s_{12}c_{13} && s_{13} \\
-s_{12}c_{23}-c_{12}s_{23}s_{13} && c_{12}c_{23}-s_{12}s_{23}s_{13} && 
s_{23}c_{13} \\
s_{12}s_{23}-c_{12}c_{23}s_{13}  && -c_{12}s_{23}-s_{12}c_{23}s_{13}&& 
c_{23}c_{13}\\
\end{array} \right)
\label{eq:2}
\end{equation}
with $c_{ij} = \cos \theta_{ij}$ and $s_{ij} = \sin \theta_{ij}$, 
where $\theta_{12}, \theta_{13}$ and  $\theta_{23}$  are three 
independent real angles lying in the first quadrant.

Of the three $\delta m_{ij}^{2}$'s appearing in eq.~(\ref{eq:1}), 
only two are independent. The complete solution of the problem 
consists therefore in determining five unknowns: two 
$\delta m_{ij}^{2}$'s and the three $\theta_{ij}$'s.

For $m_{1}\ll m_{2}\ll m_{3}$, only two $\delta m_{ij}^{2}$'s 
characterize the oscillatory behaviour of eq.~(\ref{eq:1}): 
$\delta m_{12}^{2}$  and $\delta m_{13}^{2}\simeq\delta 
m_{23}^{2}\gg\delta m_{12}^{2}$, 
corresponding, respectively, to a ``slow" and a ``fast" 
oscillation. Thus, for any given experimental situation, 
depending on the range of $L/E$ under study, either only the 
fast or both the fast and slow oscillations may be occurring.

This allows to define the two ranges $1/\delta m_{13}^{2} 
({\rm{eV}}^{-2}) \ll L/E ({\rm{m/MeV}}) \ll 1/\delta 
m_{12}^{2} ({\rm{eV}}^{-2})$  ("short-baseline") 
and $L/E ({\rm{m/MeV}}) \gg 
1/\delta m_{12}^{2} ({\rm{eV}}^{-2})$ ("long-baseline") 
in which the average transition probabilities 
$P_{\alpha\beta}^{S}$ and $P_{\alpha\beta}^{L}$ are 
calculated from eq.~(\ref{eq:1}) for 
$\sin^{2}(\Delta_{12}/2)=0$, $\langle \sin^{2}(\Delta_{13}/2)
\rangle =0.5$ and $\langle \sin^{2}(\Delta_{12}/2)\rangle=0.5$, 
$\langle \sin^{2}(\Delta_{13}/2)\rangle =0.5$, respectively.

\section{Input data} 

\subsection{Accelerator neutrinos}       
In prompt neutrinos from charm decay, the equality between the 
$\nu_{e}$ and $\nu_{\mu}$ spectra prescribed by e-$\mu$ universality 
appears to be violated as the neutrino flux asymmetry 
\begin{equation}
A=(\nu_{\mu}{~\rm{flux}}-\nu_{e}{~\rm{flux}})/
        (\nu_{\mu}{~\rm{flux}}+\nu_{e}{~\rm{flux}})   \label{eq:3}
\end{equation}
is experimentally determined to be $A=0.21 \pm 0.05$ 
\cite{GCon90}.
 
For values of $L/E$ smaller than about 0.1 m/MeV, and consequently 
in the domain of the prompt neutrino data, $P_{e\mu}$ and 
$P_{\mu\tau}$ are experimentally known to obey the 
90\% CL upper limits \cite{PDG94}
\begin{equation}
P_{e\mu}<1.5\times10^{-3}\;, \;\;\;\; P_{\mu\tau}<2\times10^{-3}\;. 
     \label{eq:4}
\end{equation}

Thus, ascribing the non-vanishing value of $A$ to a depletion of 
the $\nu_{e}$ flux due to oscillations, the analysis of the $L/E$ 
dependence of the data leads to the existence of 
$\nu_{e}$-$\nu_{\tau}$ oscillations with the parameters
\cite{GCon90,GCon92}
\begin{equation}
\sin^{2}(2\alpha)=0.48 \pm 0.10 \pm 0.05           \label{eq:5}
\end{equation}
\begin{equation}                           
\delta{m}^{2}=(377 \pm 27 \pm 7)\; \rm{eV}^{2}.  \label{eq:6}
\end{equation}

In terms of three-flavour notations, eq.~(\ref{eq:6}) ensures 
this to be the fast oscillation ($\sin^{2}(\Delta_{12}/2)=0$), 
and the limits of eq.~(\ref{eq:4}) imply that $\theta_{23}$
must be very small. It then follows that eqs.~(\ref{eq:5}) 
and (\ref{eq:6}) can be rewritten as
\begin{equation}
\sin^{2}(2\theta_{13})=0.48 \pm 0.12      \label{eq:7}
\end{equation}
\begin{equation}                    
\delta{m}_{13}^{2}\simeq\delta{m}_{23}^{2}=
               (377 \pm 29)\; \rm{eV}^{2}.      \label{eq:8}
\end{equation}

\subsection{Atmospheric neutrinos}
Neutrinos are also produced in the interactions 
of the primary component of cosmic rays in the Earth's atmosphere 
and in the subsequent decays of the produced secondaries. As 
almost all these decays involve a muon, which in turn also decays,
the ratio $R$ between the $\nu_{\mu}$ and $\nu_{e}$ fluxes can be 
safely predicted to be approximately 2. 

Several experiments have addressed this question \cite{BBar95}. 
They all consistently find values of $R$ which are smaller than 2 by 
a factor of $r$, approximately equal to 0.6. In fact, for average 
neutrino energies below 1 GeV, the weighted average of all 
available data \cite{Nus89,Frej90,IMB92,Kam92,Soud95} 
listed in Table 1 gives
\begin{equation}
r=0.63 \pm 0.06  \;\;\;\;\;  {\rm with} \;\chi^{2}/{\rm{d.o.f.}}=0.96.
\label{eq:9}
\end{equation}

\begin{table}
\caption{Atmospheric neutrino experimental results.}
\begin{center}
\begin{tabular}{lc}                         
EXPERIMENT & $r = (\mu/e)_{\rm{obs}}/(\mu/e)_{\rm{calc}}$ \\
\hline
  Nusex                     &      $ 1.04 \pm 0.32 $ \\
  Frejus (contained events) &      $ 0.87 \pm 0.19 $ \\
  IMB-3                     &      $ 0.54 \pm 0.13 $ \\
  Kamiokande                &      $ 0.60 \pm 0.08 $ \\
  Soudan 2                  &      $ 0.64 \pm 0.19 $ \\
\end{tabular}
\end{center}
\end{table}

Contrary to the low-energy behaviour, in the multi-GeV energy 
range $r$ is observed to be a function of the 
zenith-angle\cite{YFuk94}. This is very suggestive of an 
$L/E$ dependence of $r$ and thus of 
an oscillation-induced phenomenon. The $\delta{m}^{2}$ 
characterizing this oscillation lies in the 90\% CL interval
\begin{equation}
5 \times 10^{-3} < \delta{m}_{12}^{2} < 8 \times 10^{-2}  
~{\rm{eV}}^{2}.                    \label{eq:10}
\end{equation}

Eq.~(\ref{eq:10}) ensures that in the sub-GeV energy range 
all detected observables have their average long-baseline values. 
Thus, in terms of the three-flavour formalism, eq.~(\ref{eq:9}) 
becomes
\begin{equation}
(P_{\mu\mu}^{L}+\rho{P_{e\mu}^L})
(\rho^{-1}P_{e\mu}^L+P_{ee}^L)^{-1}=
0.63 \pm 0.06                     \label{eq:11}
\end{equation} 
where $\rho=0.47 \pm 0.02$ is the expected $\nu_e/\nu_{\mu}$ 
flux ratio in the absence of oscillations \cite{MHon90,BGS89}

\subsection{Solar neutrinos}
Solar neutrinos have been known for some time to deviate from 
expectations. The experimental results 
\cite{Clor95,Kam95,SAGE95,Gall95}, coming from 
three different types of detectors, as well as the predictions 
of the Princeton-Yale \cite{BP92} and Saclay \cite{TL93} 
solar models are summarized in table 2. With the only exception of 
${\rm{Ga}}^{71}$, experimental results are dominated by systematic 
errors. Furthemore, it can be seen that theoretical uncertanties 
almost always exceed experimental errors. Thus, in any comparison 
between models and experiments, theoretical errors must necessarily 
be taken into account. As they are highly correlated, we have limited 
our choice of theoretical models to those published and for which 
the complete error correlation matrix is available \cite{FL95}.

\begin{table}
\caption{Solar neutrino experimental results and theoretical 
predictions ($1\sigma$). Statistical and systematic errors have been
added in quadrature. 
For the ${\rm{Ga}}^{71}$ entry, the quoted 
experimental value is the weighted
average of the SAGE  ($ 69 \pm 10^{+5}_{-7}$  SNU 
\protect\cite{SAGE95}) 
and Gallex  ($77.1 \pm 8.5^{+4.4}_{-5.4}$ SNU 
\protect\cite{Gall95}) 
results. 
The error on this value is almost certainly underestimated as 
systematic errors have been treated as uncorrelated.}
\begin{center}
\begin{tabular}{lccc}
&\multicolumn{3}{c}{DETECTOR}\\
\hline
&${\rm{Cl}}^{37}$ &     Kamiokande                &  ${\rm{Ga}}^{71}$\\
&(SNU)             & $(10^{6}\,{\rm cm}^{-2} \rm{s}^{-1}$) & (SNU) \\
\hline
Experimental    result& 2.55$\pm$0.25           & 2.75$\pm$0.45
& 73.7$\pm$7.6 \\
Princeton/Yale model& 8$\pm$1                   & 5.7$\pm$0.8
& 132$\pm$6 \\
Saclay  model   & 6.4$\pm$1.4                 & 4.4$\pm$1.1
& 123$\pm$7 \\
\end{tabular}
\end{center}
\end{table}

With neutrino mass differences as large as those of eqs.~(\ref{eq:8}) 
and (\ref{eq:10}), all solar neutrino oscillation-induced phenomena 
observable on Earth are bound to be energy-independent. In particular, 
in all experiments the $\nu_{e}$ flux must be reduced by the same 
factor $F$, independently of the detection threshold.

Although marginally in some cases, the data are consistent with this 
expectation. Fig. 1 shows the two $\chi^{2}$'s obtained by comparing 
the experimental data with the theoretical predictions of the 
Princeton-Yale (PY) and Saclay (S) solar models as a function of $F$. 
In the analytical formulation of these $\chi^{2}$'s, theoretical 
and experimental errors have been added in quadrature and 
correlations among theoretical errors have been fully 
taken into account by means of the appropriate correlation matrices. 
The $\chi^{2}_{\rm{PY}}$ and $\chi^{2}_{\rm{S}}$ minima turn out to 
be $\chi^{2}_{\rm{PY}_{\rm{min}}} = 9.48$ (CL = 0.9\%, 
equivalent to 2.6 sigma) and $\chi^{2}_{\rm{S}_{\rm{min}}} = 5.10$ 
(CL = 7.8\%, equivalent to 1.8 sigma). 
The confidence level  of the first fit is admittedly rather low, 
but not unacceptable. In view also of the goodness of the second fit 
and of the only approximate knowledge of both theoretical and 
experimental errors (particularly with regard to their Gaussian-like 
behaviour), no conclusion about a compelling energy-dependence of the 
$F$-factor can be drawn at this stage.

\begin{figure}
\begin{center}\mbox{\epsfig{file=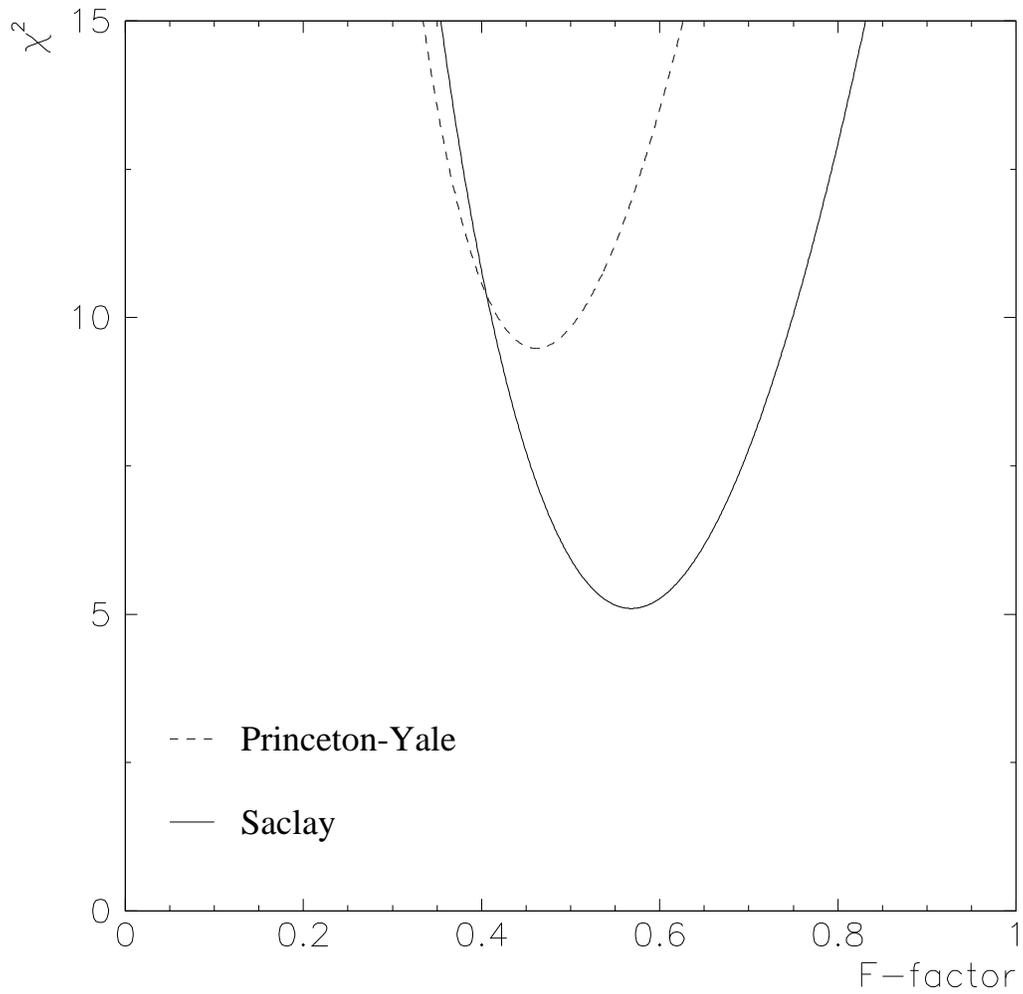,height=15cm,width=15cm}}\end{center}
\caption{The functions $\chi^2 = \chi^2(F)$ obtained by 
comparing the experimental data with the predictions of the 
Princeton-Yale (dashed line) and Saclay (full line) solar models 
(see Table 2). $F$ is the energy-independent factor by which all 
$\nu_e$ fluxes predicted by the same model are reduced. For each 
model the correlations among errors on the expected rates for the 
various detectors have been taken into account through the use 
of the appropriate error correlation matrix. All errors have 
been assumed to be Gaussian-distributed.}
\end{figure}

Following the procedure normally used by the Particle Data Group 
\cite{PDG94}, the errors on the $F$-factors have been multiplied by 
the appropriate scale-factors. Then, the two $F$-factors turn out to 
be $F_{\rm{PY}} = 0.46 \pm 0.13$ and  
$F_{\rm{S}} = 0.57 \pm 0.13$, implying the existence of a 
solar neutrino deficit. 

In order to quantify this deficit for our further analysis, we have 
calculated a new $F$-factor taking the arithmetic average of 
$F_{\rm{PY}}$ and $F_{\rm{S}}$. We have taken the 
error on this average ($\pm 0.08$) to be a good estimator of a 
systematic error reflecting the fact that the two models of the 
same Sun yield different predictions. We have added this error in 
quadrature to the typical statistical error ($\pm 0.13$), arriving 
at the result
\begin{equation}
F = 0.51 \pm 0.15.  \label{eq:12}
\end{equation}

Using the formalism of the three-flavour analysis, eq.~(\ref{eq:12}) 
trivially translates into
\begin{equation}
P_{ee}^{L} = 0.51 \pm 0.15.  \label{eq:13}
\end{equation}

\section{Mass eigenstates and mixing matrix}
        
Taking $m_{1} \ll m_{2}$,  eqs.~(\ref{eq:8}) and 
(\ref{eq:10}) yield
\begin{equation}
m_{1} \ll 10^{-2}~{\rm{eV}}\; , \;\;\;\;  m_{2} = (0.18 \pm 0.06) 
~{\rm{eV}}\;,  \;\;\;\; m_{3} = (19.4 \pm 0.7)~{\rm{eV}}\;  
\label{eq:14}
\end{equation}

These values are consistent with the predictions of the see-saw 
model \cite{GMRS}. Using a leptonic quadratic hierarchy and the above 
value of $m_3$ one has $m_1=1.6\times10^{-6}$ eV and $m_2=6.8
\times10^{-2}$ eV.
        
From the three eqs.~(\ref{eq:7}), (\ref{eq:11}) and (\ref{eq:13}), 
within the constraints~(\ref{eq:4}), 
the three angles $\theta_{ij}$ are uniquely determined to be
\begin{equation}
\theta_{12}=0.55 \pm 0.08\; , \;\;\;\;  \theta_{13}=0.38\pm 0.06\;,
\;\;\;\; \theta_{23} < 0.03\;.
\label{eq:15}
\end{equation}
with $\chi^{2}_{\rm{min}} = 0.093$.

As in any solution with $m_3$ larger than a few eV, the smallness of 
$\theta_{23}$ is dictated by the upper limits 
(\ref{eq:4}) and this implies an 
over-determination of the angles $\theta_{12}$ and $\theta_{13}$. 
The good consistency of the 
input data is illustrated in fig. 2, which shows the three relations 
between $\theta_{12}$ and $\theta_{13}$ obtained from 
eqs.~(\ref{eq:7}), (\ref{eq:11}) and (\ref{eq:13}) for 
$\theta_{23} = 0$. It can be seen, for instance, that the value of 
$\theta_{13}$ is determined not just by the result of 
eq.~(\ref{eq:7}) but also by the system formed by the other two.

\begin{figure}
\begin{center}\mbox{\epsfig{file=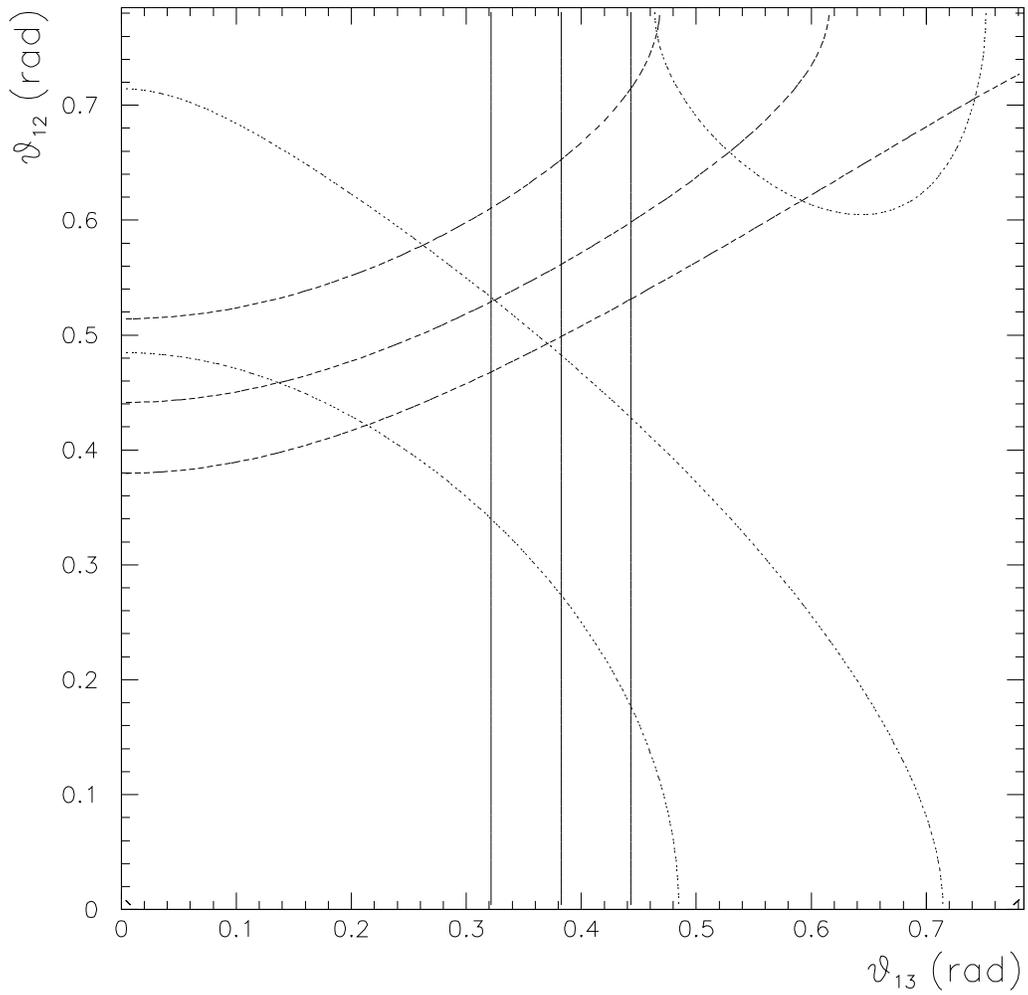,height=15cm,width=15cm}}\end{center}
\caption{
The three relations ($\pm 1\sigma$) between $\theta_{12}$ 
and $\theta_{13}$ obtained from eqs.~(\protect\ref{eq:7}) 
(full lines), (\protect\ref{eq:11}) 
(dashed lines) and (\protect\ref{eq:13}) (dotted lines) 
for $\theta_{23}=0$.
}
\end{figure}

The $U$-matrix is then
\begin{equation}
U = \left( \begin{array}{ccccc}
0.79 \pm 0.05  && 0.49 \pm 0.06 &&  0.37 \pm 0.05  \\
-0.52 \pm 0.06  && 0.85 \pm 0.05  && < 0.03   \\
-0.31 \pm 0.04  && -0.20 \pm 0.05  && 0.93 \pm 0.03
\end{array} \right)
\label{eq:16}
\end{equation}

The knowledge of the angles $\theta_{ij}$ allows also to calculate 
the following short- and long-baseline average transition 
probabilities $P_{\alpha\beta}^{S}$ and $P_{\alpha\beta}^{L}$ 
\[
\begin{array}{lclcl}   
P_{ee}^{S} = 0.77\pm 0.06  && P_{e\mu}^{S} < 3\times 10^{-4} 
&& P_{e\tau}^{S} = 0.23\pm 0.06 \\
                           && P_{\mu\mu}^{S} > 0.998         
&& P_{\mu\tau}^{S} < 1.5\times 10^{-3} \\
                         &&                                
&& P_{\tau\tau}^{S} = 0.77\pm 0.06
\end{array}
\]
\[
\begin{array}{lclcl}   
P_{ee}^{L} = 0.47 \pm 0.06   && P_{e\mu}^{L} = 0.35 \pm 0.05   
&& P_{e\tau}^{L} = 0.19 \pm 0.05 \\ 
                           && P_{\mu\mu}^{L} = 0.60 \pm 0.06
&& P_{\mu\tau}^{L} = 0.05 \pm 0.03 \\
                           &&                                    
&& P_{\tau\tau}^{L} = 0.76\pm 0.06
\end{array}
\]

All upper limits are at the 90\% CL.

\section{Discussion}

The results for the mixing angles (eq.~(\ref{eq:15})) 
are that $\theta_{12}$ is fairly sizeable, $\theta_{23}$ 
is small and $\theta_{13}$ lies somewhere 
in between the two. Small values of $\theta_{23}$ are 
typical of any solution in which the mass of the 
heaviest eigenstate (eq.~(\ref{eq:14})) is of  
cosmological relevance \cite{DeRu80}.
        
The main implications of this scenario are discussed below.

\subsection{Accelerators}
Short-baseline experiments like Chorus and Nomad \cite{KWin95} 
are expected to see a $\nu_{\tau}$ signal. If not from 
$\nu_{\mu}$-$\nu_{\tau}$
($P_{\mu\tau}^S$  is consistent with zero), 
at least from $\nu_{e}$-$\nu_{\tau}$ transitions 
(owing to the about 1\% $\nu_{e}$ component in the beam 
and the sizeable $P_{e\tau}^S$). 

Long-baseline experiments \cite{BBar95,JSch95} have a better chance 
to detect $\nu_{\mu}$-$\nu_{\tau}$ oscillations, but the largest 
effect is anticipated in the $\nu_{e}$-$\nu_{\mu}$ channel. 

Experiments at low energy accelerators like KARMEN \cite{KARM95} and 
LSND \cite{LSND95} 
ought to be able to detect $\nu_{e}$-$\nu_{\mu}$ transitions as, 
owing to the onset of the slow oscillation, $P_{e\mu}$ in their 
accepted range of $L/E$ is typically a fraction of a percent. 
Both experiments should be in the position of investigating the 
conservation of the $\nu_e$  flux, a study which 
so far has resulted in the KARMEN not yet statistically significant 
$P_{ee}^S$ lower limit \cite{KARM95}.

\subsection{Beta decay and reactors}
For Majorana neutrinos, under very reasonable assumptions, the 
neutrino-less double beta decay amplitude is proportional 
to \cite{Moe95}
\begin{equation}
m_{\beta\beta}=\sum_{i}U_{ei}^{2}m_i
\label{eq:17}
\end{equation}
The value of $m_{\beta\beta}$ calculated from eqs. (\ref{eq:14}) and
(\ref{eq:16}) is 2.6 eV, while its 
present upper limit is only about several eV \cite{PDG94}, so that 
experiments will 
soon be in the position of providing definite conclusions on this very 
important issue.

In about 13\% of all $\beta$-decays a heavy neutrino of mass $m_3$
(eq.(\ref{eq:14})) is expected to be produced. Two-state analyses 
of the spectra end-points \cite{Bele95} might be able to investigate 
this question. However, at the present time, the presence of 
ununderstood phenomena in the high energy region precludes 
any firm conclusions \cite{Wark95}.
        
All $\beta$-sources are expected to yield 
lower-than-canonically-calculated 
$\nu_e$ fluxes (both $P_{ee}^S$ and $P_{ee}^L$ are 
smaller than 1 and the wave-length of the 
fast oscillation is about 5 mm/MeV) but the only result available 
so far \cite{Bach95} is not accurate enough to really test this 
possibility.
        
This effect should also be present in experiments at nuclear reactors.
After some initial tantalizing results indicating a large depletion 
of the $\nu_{e}$  flux at the two sigma level \cite{Rein83,Kwon81}, 
the more recent experiments have failed to substantiate any deviation 
from expectations. In fact, their claims represent the most stringent 
limit to oscillations of the $\nu_{e}$ and the most serious challenge 
to the result of eq. (\ref{eq:5}).
        
For mass differences as large as that of eq. (\ref{eq:6}), owing to 
the very short wavelength of the oscillation, these experiments  
consist in comparing the antineutrino flux expected from the reactor 
with that experimentally measured at some (short-baseline) distance 
away. In spite of the many such experiments carried out in the last 
decade \cite{Goesg86,Rovn91,Krasn94,BuKu94,Bug95}, no significant 
progress has been made in the accuracy of the results. The comparison 
of ref. \cite{Goesg86} with ref. \cite{Bug95} (the two best 
documented and, respectively, the oldest and newest papers) shows 
that in both cases the determination of the ratio between results 
and expectations is limited by a systematic error of about 6\%.
This error is largely dominated by a common uncertainty on the 
knowledge of the reactor antineutrino energy spectrum. This is due to 
the fact that the bulk of the data on thermal neutron fission induced 
$\beta$-spectra \cite{Pu239,U235,Pu241} all reactor antineutrino 
energy spectra calculations are based on were obtained 
only once, more than a decade ago. It has been argued repeatedly that 
such important measurements ought to be checked, that the procedure 
used in deriving neutrino spectra from $\beta$-spectra ought to be 
better understood, that the quoted systematic uncertainty is probably 
underestimated and that the over-all normalization error in the 
reactor antineutrino energy spectrum is more likely to be around 
10\% \cite{Bouch86,Ket88,Afo88}. In any case, even taking the quoted 
6\% at face value and assuming it to be Gaussian-distributed, the 
reactor result differs from the value of $P_{ee}^S$ resulting from eq. 
(\ref{eq:15}) by 2.7 sigma, indicating some discrepancy but not a 
really serious incompatibility.
        
Reactor experiments are potentially sensitive also to the existence 
of the mass difference of eq. (\ref{eq:10}). In this case the 
wavelength of the oscillation is certainly long enough to allow a 
comparison between two measurements of the antineutrino flux in two 
detectors at different distances from the reactor, thus reducing the 
systematic effects discussed above. Because of the low event rate, 
experiments so far have been barely able to investigate the $L/E$ 
region of interest. However, the claimed limits 
\cite{Goesg86,Rovn91,Krasn94,BuKu94,Bug95} 
exclude in part or almost completely the region 
of eq. (\ref{eq:10}). On the other hand, even if in a different 
context, the re-analysis of the reactor data of ref. \cite{HPS95} has 
shown that a mass $m = (0.085 \pm 0.010)$ eV is quite consistent with 
all existing evidence. Reactor experiments with an even slightly 
larger $L$ than it has been possible so far are clearly needed 
to clarify this important point. Long-base experiments will also 
have the benefit of a projected larger depletion of the $\nu_e$ flux 
($P_{ee}^L < P_{ee}^S$).
        
In conclusion, although reactor experiments do indicate the existence 
of some potential problems at the very edges of the explored region 
in the $\sin^{2}(2\alpha),{\delta}m^{2}$ plane, the large 
uncertainties involved preclude any firm conclusion at the present 
time.

\subsection{Astrophysics and cosmology}
Neutrinos with masses of the order of ten eV can be the dark matter 
particles needed to explain the high rotation velocities in the outer 
parts of spiral galaxies \cite{Ingr92,Moo94} and the observed X-ray 
emission from hot diffused gas in elliptical galaxies \cite{DeP95}. 
        
Heavy neutrinos lead also to some important cosmological consequences.
With masses as large as those of eq. (\ref{eq:14}), neutrinos 
contribute far more than baryons to the total matter density of the 
universe $\Omega_{0} = \rho_{0}/\rho_{c}$ where $\rho_{c}$ 
is the critical density (the subscript $0$ indicates present-day 
values).
        
In terms of the Hubble parameter $H_0\; (H_{0} = 100\, h_{0}\, 
{\rm{km}}\, {\rm{s}}^{-1}\, {\rm{Mpc}}^{-1}$) the 
baryon density $\Omega_b$ is known to be \cite{PDG94}
\begin{equation}
\Omega_{b}h_{0}^{2} = (0.015 \pm 0.005)
\label{eq:18}
\end{equation}
while the neutrino density $\Omega_{\nu}$ obtained from eq. 
(\ref{eq:14}) is
\begin{equation}
\Omega_{\nu}h_{0}^2 = 1.075\times10^{-2}\times \sum_i{m}_i = 
(0.211 \pm 0.008).
\label{eq:19}
\end{equation}

Thus, for a universe whose only massive stable components are 
baryons and neutrinos, the total density is given by
\begin{equation}
\Omega_{0}h_{0}^2 = (\Omega_b+\Omega_{\nu})h_{0}^2 = (0.226 \pm 0.009)
\label{eq:20}
\end{equation}

The relation resulting from eq. (\ref{eq:20}) is shown in the 
$\Omega_0,H_0$ plane of Fig. 3 by the full-line curves. They define 
the allowed region (1$\sigma$) for a universe composed only of 
baryons (visible matter) and neutrinos (hot dark matter, HDM). The 
allowed region for a possible third massive component (cold dark 
matter, CDM) lies on the right-hand side of these 
curves. The same figure shows also the relations between 
$\Omega_0$ and $H_0$ obtained from the Einstein's equations for the 
two limiting values of the age 
of the universe $T=13,16$ Gyr and for two different values of the 
cosmological constant $\Lambda$ ($\Lambda=3H_0^2\lambda_0/c^2$) 
$\lambda_0=0,1$.

\begin{figure}
\begin{center}\mbox{\epsfig{file=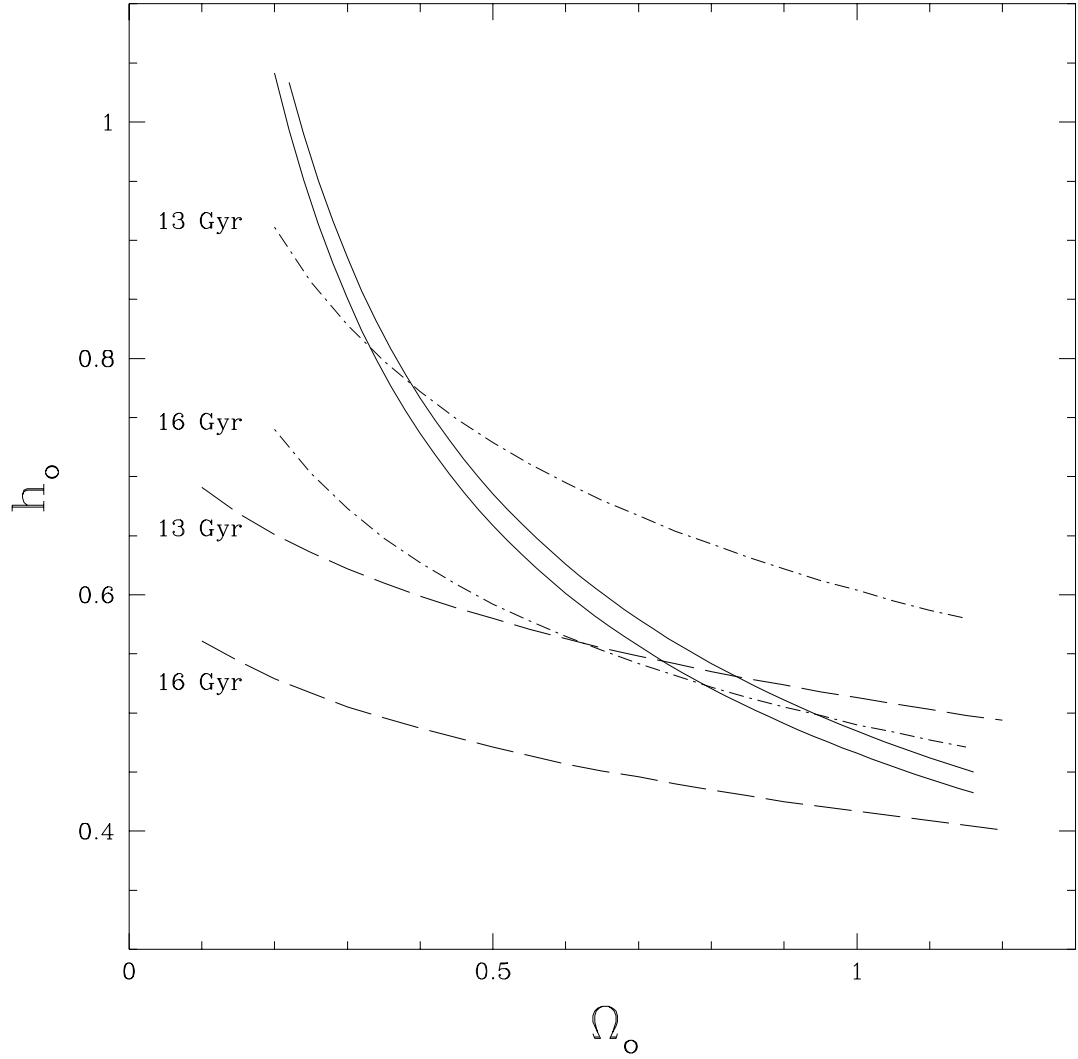,height=15cm,width=15cm}}\end{center}
\caption{Relations between the total matter density of the universe 
$\Omega_0$ (in units of critical density) and the normalized Hubble 
parameter 
$h_0$ ($H_0=100\,h_0\, {\rm km~{s}^{-1} {Mpc}^{-1}}$). 
The full lines are for a sum of the neutrino masses 
of ($19.6 \pm 0.7$) eV. The other curves represent the function 
$h_0 = h_0(\Omega_0)$ 
for the two limiting values of the age of the universe $T=13$ Gyr and 
$T=16$ Gyr 
and for values of the scaled cosmological constant $\lambda_0$ 
($\lambda_0=\Lambda{c}^2/3H_0^2$) $\lambda_0=0$ 
(dashed lines) and $\lambda_0=1$ (dotted-dashed lines).}
\end{figure}
        
To date, the value of $H_0$ is not precisely known, the various 
experimental determinations ranging between the values 
$h_0\simeq 0.5$ and $h_0\simeq 0.8$ \cite{H0}. 
        
Depending on the value of $H_0$, different situations can occur.
\begin{enumerate}
\item For $h_0\simeq 0.5$ the choice $\lambda_0 = 0$ is consistent 
with the accepted bounds on the age of the universe. In this case 
eq. (\ref{eq:20}) requires $\Omega_0 \gtrsim 0.85$ and for 
$\Omega_0 = 1$ neutrinos turn out to be the most massive component of 
the universe. For only baryons and neutrinos 
contributing to the total density of a critical universe, the result 
of eq. (\ref{eq:20}) implies
\begin{eqnarray}
&H_{0}& = (47.4 \pm 0.9)\, {\rm{km}} \, {\rm{s}}^{-1}\, {\rm{Mpc}}^{-1}
\nonumber \\
&& \label{eq:21} \\
&T& = (13.8\pm 0.3)\, {\rm{Gyr}}. \nonumber
\end{eqnarray}

\item For $h_0 \simeq 0.65$, eq. (\ref{eq:20}) requires 
$\lambda_0 > 0$. 
For $\lambda_0 = 1$,  
$\Omega_0 \gtrsim 0.5$  and the presence of cold dark matter could be 
accomodated with a ratio $\Omega_{CDM}/\Omega_{HDM}\,$ 
($\Omega_{CDM}/\Omega_{\nu}$) 
which, depending on the value of $\Omega_0$, 
could be as large as $\approx 0.5$.

\item For $h_0 \simeq 0.8$ and $\lambda_0 = 1$, $\Omega_0 \simeq 0.4$ 
and the mass of the universe is again due solely to baryons and 
neutrinos. Non-vanishing values of $\Omega_{CDM}/\Omega_{HDM}$  are 
possible only for $\lambda_0 > 1$.
\end{enumerate}
        
Thus, for any reasonable values of $H_0$  and $\lambda_0$, neutrinos 
with masses as large as those of eq. (\ref{eq:14}) are the dominant 
component of the mass of the universe. As a consequence, their 
existence contradicts all cold-dark-matter-dominated (CHDM) 
cosmological models \cite{Pri95}. To have the same success that 
these models have in explaining the present data on the 
power spectrum of density fluctuations, hot-dark-matter-dominated 
models will have to rely on some new hypotheses, such as, for 
instance, seeds of non-Gaussian origin \cite{Alb92,Turn95}.

\section{Conclusions}

This analysis shows that, at least taking results at face value, 
the evidence for the existence of neutrino oscillations is really 
substantial. Although none of the existing positive signals is 
admittedly beyond some criticism, their ensemble is quite compelling.
        
The signature of neutrino oscillations lies in the $\sin^2$ factor of 
eq. (\ref{eq:1}) and any convincing proof of their existence must 
rely on some experimental observation of an $L/E$ dependence. This 
evidence, first obtained from the analysis of the $L/E$ modulation 
of the $\nu_{\mu}$-$\nu_e$ asymmetry in prompt neutrinos 
\cite{GCon90,GCon92}, has been further strengthened recently 
by the observation of the dependence on the zenith-angle (effectively 
also $L/E$) of the ratio between the $\nu_{\mu}$ and $\nu_{e}$ fluxes 
in atmospheric neutrinos of multi-GeV energy
\cite{YFuk94}.
        
All the available experimental information from accelerator, 
atmospheric and solar neutrinos is accounted for in the framework of a 
three-flavour neutrino oscillation analysis. The solution found is 
unique and not significantly contradicted by any existing result, all 
conflicting evidence being below the three sigma level.
        
Neutrino physics is presently at a crossroads. If $m_3$ is indeed 
large enough to be of cosmological relevance, the limits of eq. 
(\ref{eq:4}) imply that $\theta_{23}$  is very small. The relations  
between $\theta_{12}$ and $\theta_{13}$ are then those 
of fig. 2 and all indications are that $\theta_{13}$  is sizeable. 
Under this condition, high energy accelerator experiments such as 
Chorus and Nomad should not fail to see a $\nu_{\tau}$ signal. If 
they do, this implies that at least some of the input data to our 
analysis are wrong and that either $\theta_{13}$ or $m_3$ or both are 
smaller than the results we have arrived at. In the first case, the 
atmospheric neutrino result (see fig. 2) implies the relative large 
value for the $F$-factor $F=0.70 \pm 0.07$, thus disfavouring 
solar models such as that of Princeton-Yale \cite{BP92} which 
(see fig. 1) require $F \lesssim 0.5$. In the other two cases 
$m_3$ must be smaller than a few eV, thus excluding neutrinos 
as the main constituents of the dark matter of the universe and 
leaving them out of a primary role in astrophysics.

\end{document}